\documentclass[
aps,
prc,
showpacs,
twocolumn,
superscriptaddress,
nofootinbib,
floatfix]
{revtex4}
\usepackage{graphicx,
longtable,
slashed}%

\begin{document}  

\title{Evaluation of the resonance enhancement effect in neutrinoless double-electron capture in $^{152}$Gd, $^{164}$Er and $^{180}$W atoms}

\author{Dong-Liang Fang}  
\affiliation{Institut f\"{u}r Theoretische Physik der Universit\"{a}t  
T\"{u}bingen, D-72076 T\"{u}bingen, Germany} 
\author{K. Blaum} 
\affiliation{Max-Planck-Institut f\"ur Kernphysik, Saupfercheckweg 1, D-69117 Heidelberg, Germany} 
\affiliation{Fakult\"at f\"ur Physik und Astronomie, Ruprecht-Karls-Universit\"at Heidelberg, D-69117 Heidelberg, Germany} 
\author{S. Eliseev} 
\affiliation{Max-Planck-Institut f\"ur Kernphysik, Saupfercheckweg 1, D-69117 Heidelberg, Germany} 
\author{Amand Faessler} 
\affiliation{Institut f\"{u}r Theoretische Physik der Universit\"{a}t  
T\"{u}bingen, D-72076 T\"{u}bingen, Germany} 
\author{M.I. Krivoruchenko} 
\affiliation{Institute for Theoretical and Experimental Physics, B. Cheremushkinskaya 25 \\ 
117218 Moscow, Russia} 
\affiliation{Department of Nano-, Bio-, Information and Cognitive Technologies\\ 
Moscow Institute of Physics and Technology, 9 Institutskii per. \\ 
141700 Dolgoprudny, Moscow Region, Russia} 
\author{V. Rodin} 
\affiliation{Institut f\"{u}r Theoretische Physik der Universit\"{a}t  
T\"{u}bingen, D-72076 T\"{u}bingen, Germany} 
\author{F. \v Simkovic} 
\affiliation{Department of Nuclear Physics and Biophysics,  
Comenius University, Mlynsk\'a dolina F1, SK-84215 
Bratislava, Slovakia} 
\affiliation{Laboratory of Theoretical Physics, JINR, 
RU-141980 Dubna, Moscow region, Russia} 
 
\date{\today}

\begin{abstract} 
We study the resonant neutrinoless double-electron capture  
($0\nu$ECEC) in $^{152}\mathrm{Gd}$, $^{164}\mathrm{Er}$ and $^{180}\mathrm{W}$ atoms, 
associated with the ground-state to ground-state nuclear transitions. The corresponding  
matrix elements are calculated within the deformed QRPA using the realistic Bonn-CD nucleon-nucleon interaction. 
The half-lives are estimated with the use of the most recent precision data on the $Q$-values of these processes. 
Perspectives of experimental search for the $0\nu$ECEC with the isotopes $^{152}\mathrm{Gd}$, $^{164}\mathrm{Er}$ 
and $^{180}\mathrm{W}$ are discussed.  
\end{abstract}

 
 
\maketitle 
 
\section{Introduction} 
 
One of the unsolved mysteries of today's particle physics and cosmology  
is the question of whether neutrinos are Dirac or Majorana particles.
In the first case neutrinos and antineutrinos are fundamentally different,  
whereas in the second case neutrinos and antineutrinos are identical.  
Theoretical arguments in favor of Majorana neutrinos exist for decades  
in connection with the smallness of neutrino masses. Some grand unified theories  
explain smallness of the masses, e.g., in a seesaw scenario with  
heavy Majorana leptons \cite{seesaw}. 
 
Neutrinoless double-beta decay ($0\nu\beta\beta$),  
\begin{equation} 
(A,Z) \rightarrow (A,Z+2) + e^- + e^-, 
\label{e:1} 
\end{equation} 
is being considered as the unique practical tool for determining the nature of neutrinos  
(for a review, see \cite{dbdrev}). 
The most favorable decays for the experimental search 
are those with high mass difference between the ground state neutral atoms, i.e. $Q$-values, 
in which the parent nuclei decay to ground states of the daughter nuclei. 
 
The mere observation of neutrinoless double-electron capture ($0\nu$ECEC),  
\begin{equation} 
e^-_b + e^-_b + (A,Z) \rightarrow (A,Z-2)^{**},  
\label{e:2} 
\end{equation} 
could also prove the Majorana nature of neutrinos as well as the violation of the total lepton number conservation.  
A double asterisk in Eq.~(\ref{e:2}) means that, in general, the final atom $(A,Z-2)$ is excited with respect  
to both the electron shell, due to formation of two vacancies for the electrons, and the nucleus. This process,  
as noted long time ago by Bernab\'eu, De Rujula and Jarlskog \cite{DERU}, may have a resonant character under the condition  
of degeneracy of the masses of initial and intermediate atoms. In contrast to the $0\nu\beta\beta$ decays, in the $0\nu$ECEC  
capture small $Q$-values are favorable. The capture rate is a sensitive measure of the neutrino mass. 
 
The resonance enhancement can increase the probability of capture by many orders of magnitude. In searches  
for Majorana neutrinos, the neutrinoless double-electron capture can compete with the neutrinoless double-$\beta$ decay  
provided the resonance condition is satisfied within a few tens of electron-volts. So far, however, there was no way to  
identify promising isotopes for experimental search of $ 0\nu$ECEC, 
because of poor experimental accuracy of measurement of $Q$-values which until recently were known with uncertainties 
of 1 - 10 keV only \cite{AME2003}. 
Progress in precision measurement of atomic masses with Penning traps \cite{penning1,penning2,penning3} 
has revived the interest in the old idea on the resonance $0\nu$ECEC capture. 
 
Sujkowski and Wycech \cite{sujwy} and Lukaszuk et al. \cite{lukas} analyzed the $0\nu$ECEC process for  nuclear $0^+ \to 0^+$ transitions accompanied by a photon emission in the resonance and non-resonance modes. The physical background for the  
process was calculated.  
A new theoretical approach developed by \v Simkovic and Krivoruchenko \cite{SIM08} and 
Krivoruchenko  et al. \cite{mik11} allowed a unified description of the oscillations of  
stable and quasistationary atoms, which take place with violation of the total lepton number conservation and are followed  
by de-excitation with emission of photons. Based on the most recent data and realistic evaluation of the decay half-lives,  a complete list  
of the most perspective isotopes for which the $0\nu$ECEC capture may have the resonance enhancement was provided in Ref.~\cite{mik11} 
for further experimental study. Some isotopes such as $^{156}$Dy have several closely-lying resonance levels. 
A more accurate measurement of $Q$-value of atoms $^{156}$Dy and $^{156}$Gd confirmed the existence of overlapping $0\nu$ECEC resonance levels \cite{elis3}.
Assuming an effective mass for the Majorana neutrino of 50 meV and an appropriate value
of nuclear matrix element, half-lives of some of the isotopes were found to be as low as $10^{25}$ years in the unitary limit,
which is one order of magnitude shorter than the $0\nu\beta\beta$ half-life of $^{76}$Ge for the same mass of Majorana neutrino \cite{erice11}.  

In high-$Z$ atoms, the electrons in inner shells are moving with relativistic velocities.  
Effects associated with the relativistic structure of the electron shells reduce the 
$0\nu$ECEC half-lives by almost one order of magnitude.  
In contrast to the non-relativistic theory, the capture of electrons from the $np_{1/2}$ states 
is only moderately suppressed in comparison with the capture from the $ns_{1/2}$ states. 
In the relativistic formalism, selection rules appear to require that nuclear transitions with a change 
in the nuclear spin $ \Delta J \geq 2 $ are strongly suppressed.  
The relativistic effects also 
enhance the violation of parity in the the $0\nu$ECEC process,  
as a result of which nuclear transitions $0^+ \to 0^{\pm},1^{\pm}$ become all attainable  
for a mixed capture of $s$- and $p$-wave electrons \cite{mik11}. 
A similar effect occurs due to weak right-hand currents as discussed by Vergados \cite{verg11}. 
 
Recently there has been fast progress in the measurement of atomic masses with the help of Penning traps.  
The accuracy of $Q$-values at around 100 eV was  
achieved \cite{elis3,redshaw09,SCIE09,RAKH09,kolhinen,mount10,elis1,elis2,elis4,droese11},  
which has already allowed to exclude a number of isotopes from the list of the most promising candidates  
for searching the neutrinoless double-electron capture.  
The best candidate is currently $^{152}$Gd, which although does not reach the unitary limit,  
however, undergoes a significant increase in the capture rate due to the proximity  
to the resonance level \cite{elis1}. Further precise measurements of masses of prospective isotopes  
are vigorously encouraged to continue.

Neutrinoless double-electron capture has a number of important advantages with respect to 
experimental signatures and background conditions.  
In recent years, experimental searches for the capture process were continued 
\cite{barab1,barab2,belli09,rukh11,frek11,belli11,belli11a}. New upper limits of about  
$ 10^{19} - 10^{21} $ years for the half-lives of $ ^{74}$Se, $ ^{106} $Cd and $ ^{112} $Sn 
have been obtained \cite{barab1,barab2,rukh11,frek11,belli11a}. 
To make further progress new experimental data on excited states of finite-state nuclei  
(excitation energy, angular momentum, parity) and precise calculations of transition 
nuclear matrix elements (NMEs) are required.  
Among the promising isotopes, $^{152}$Gd, $^{164}$Er and $^{180}$W  
have likely resonance transitions to the $0^+$ ground states of the final nuclei. 
\footnote{ 
The transition of $^{106}$Cd to an excited state of $^{106}$Pd with  
the nuclear excitation energy of 2717.59 keV was examined in detail in Ref. \cite{suho2011} 
for $J_f = 0$. In Ref. \cite{mik11} it was noted that, as long as this level 
$\gamma$-decays by 100\% into the $3^+$ state at 1557.68 keV, the assignment $J_f = 0$ is excluded. 
} 
 
In this paper, accurate calculations of the 
$0\nu$ECEC half-lives of $^{152}$Gd, $^{164}$Er and $^{180}$W are performed.  
The electron wave functions in the atoms are treated in the relativistic Dirac-Hartree-Fock approximation \cite{MAWA1973}.  
The nuclear matrix elements are calculated within the proton-neutron deformed quasiparticle  random-phase approximation (deformed QRPA) with a realistic residual interaction \cite{dqrpa1,dqrpa2,dqrpa3,dqrpa4}.

\section{Resonant neutrinoless double-electron capture} 
 
The $0\nu$ECEC leads to a violation of conservation of total lepton number by two units. If the process is due to the Majorana neutrino exchange mechanism,  
the capture rate is determined by the effective Majorana neutrino mass 
\begin{equation}\label{Majmass} 
m_{\beta\beta}=\sum_{i=1}^3 U_{ei}^{2} m_{i}. 
\label{e:3} 
\end{equation} 
Here, $U_{ei}$ is the element of the Pontecorvo-Maki-Nakagawa-Sakata  
neutrino mixing matrix \cite{BPont,MNS} and $m_i$ are the diagonal Majorana 
neutrino masses. 
 
The half-life of the $0\nu$ECEC has the form  
\begin{equation} 
T^{0\nu\mathrm{ECEC}}_{1/2} = \frac{\ln{2}}{\Gamma^{0\nu\mathrm{ECEC}}},
\label{e:4} 
\end{equation} 
where the decay width is given by 
\begin{equation} 
\Gamma^{0\nu\mathrm{ECEC}}_{a b} = \frac{\left| V_{a b} \right|^2} 
{\Delta^2  
+ \frac{1}{4} \Gamma^2_{ab}} \Gamma_{ab}, 
\label{e:5} 
\end{equation} 
with   
\begin{equation} 
\Delta = M_{A,Z} - M_{A,Z - 2}^{*} = Q - B_{a b} 
\label{e:6} 
\end{equation} 
being the difference of masses of the initial and final excited atoms with masses $M_{A,Z}$ and $M_{A,Z - 2}^{*}$, 
respectively. 
$B_{ a b} = E_{a} + E_{b} + E_C $ is the energy of two electron holes, whose quantum numbers $(n, j, l)$ are denoted by indices $a$ and $b$,  
$E_C$ is the interaction energy of the two holes, while $\Gamma_{a b}$ is the width of the excited final atom with the electron holes.
Since we consider the transitions to ground states of the final nuclei,  
the single asterisk for $M_{A,Z - 2}^{*}$ refers to the excitation of electron shells only. 

\begin{table*}[htb] 
\centering 
\caption{Phenomenological paring gaps for protons $\Delta_p$ and neutrons $\Delta_n$  
and deformation parameters $\beta$  deduced from intrinsic quadrupole moments  measured 
by the Coulomb excitation reorientation technique ($\beta_{Q_p}$, sign is given explicitly if known)~\cite{stone}  
and $B(E2)$ values ($\beta_{B(E2)}$)~\cite{raman} in $^{152}$Gd, $^{164}$Er and $^{180}$W.  
$\beta_2$ is the deformation parameter of Woods-Saxon mean field fitted to reproduce the experimental 
quadrupole moments. $\langle BCS_i | BCS_f \rangle$ is the  BCS overlap between the initial  
and final BCS vacua~\cite{deform}. }\label{table.1} 
\centering 
\renewcommand{\arraystretch}{1.1}  
\begin{tabular}{lcccccc} 
\hline\hline  
Initial (final)   & $\Delta_p$ & $\Delta_n$ & $\beta_{Q_p}$ & $\beta_ {B(E2)}$ & $\beta_2$ &  $\langle BCS_i | BCS_f \rangle$ \\  
~~~nucleus & (MeV) & (MeV) & & & & \\ \hline  
${^{152}}$Gd (${^{152}}$Sm)  & 1.478 (1.117) & 1.179 (1.192)  &  \phantom{0.36} (+0.29)    & 0.212 (0.306)  & 0.166 (0.256) &  0.44 \\ 
${^{164}}$Er (${^{164}}$Dy)  & 1.025 (0.879) & 1.020 (0.825)  & 0.36 (+0.32)    & 0.333 (0.348)  & 0.289 (0.302) &  0.73 \\ 
${^{180}}$W~ (${^{180}}$Hf)  & 0.927 (0.832) & 0.788 (0.713) & 0.27 (+0.27) & 0.252 (0.273)  & 0.237 (0.244) &  0.75 \\ 
\hline 
\hline 
\end{tabular} 
\end{table*} 
 
Having factorized the electron shell structure and the nuclear matrix 
element, the lepton number violating transition amplitude can be represented as \cite{mik11} 
\begin{eqnarray} 
V_{a b} = m_{\beta\beta} G^2_{\beta} \frac{g^2_A}{4 \pi R} \langle F_{a b} \rangle M^{0\nu}. 
\label{e:7} 
\end{eqnarray} 
Here, $G_{\beta}=G_F \cos \theta_C$, where $\theta_C$ is the Cabbibo angle, 
$g_A$ is the axial-vector coupling constant,  
$R$ is the nuclear radius,  
$\langle F_{a b} \rangle $ is a combination of averaged upper and lower bispinor components 
of the atomic electron wave functions defined in Ref. \cite{mik11}. 
$M^{0\nu}$ is the nuclear matrix element of $0^+$ ground state to $0^+$ ground state transition,  
which is a sum of the Fermi (F), Gamow-Teller (GT),  
and tensor (T) contributions \cite{Sim99, anatomy}:  
\begin{eqnarray} 
{M}^{0\nu} = - \frac{M_F}{g_A^2} + M_{GT} + M_{T}. 
\label{e:8} 
\end{eqnarray} 
In comparison with the corresponding $0\nu\beta\beta$ decays,  
the isospin operators $\tau^+$ of nucleons entering the NMEs are replaced by $\tau^-$. 
 
\section{Calculation of NMEs} 
 
Nuclei participating in the $0\nu$ECEC ground state to ground state nuclear transitions   
${^{152}\mathrm{Gd}}\rightarrow {^{152}\mathrm{Sm}}$, ${^{164}\mathrm{Er}}\rightarrow {^{164}\mathrm{Dy}}$ and  
${^{180}\mathrm{W}} \rightarrow {^{180}\mathrm{Hf}}$ are deformed.  The deformation parameter  $\beta = \sqrt{\pi/5}Q_p/(Zr^2_c)$ can be deduced from the intrinsic quadrupole moment $Q_p$ of the first $2^+$ state measured by the Coulomb excitation reorientation technique 
($r_c$ is the root mean square charge radius). Unfortunately, the electric quadrupole moment of  
${^{152}\mathrm{Gd}}$ has not been measured yet by this method~\cite{stone}.
Alternatively, the deformation parameter $\beta$  can be extracted
from the values of measured E2 transition probability, ($|Q_p| = \sqrt{16\pi B(E2)/5e^2}$,  the sign cannot be extracted~\cite{raman}). From Table \ref{table.1} one can see that the deformation parameters determined in both ways agree quite well with each other.  
 
NMEs of the considered $0\nu$ECEC transitions  
are calculated within the deformed QRPA with a realistic nucleon-nucleon  
interaction \cite{dqrpa1,dqrpa2,dqrpa3,dqrpa4}.  The details of the formalism  
for the $0\nu\beta\beta$-decay NME are given in Refs.~\cite{dqrpa3,dqrpa4}.   
The needed generalization of the basic equations of Refs.~\cite{dqrpa3,dqrpa4} to the case of $0\nu$ECEC NME is straightforwardly achieved by replacing $\tau^+$ with $\tau^-$ operator. We note that in the calculation of  
$M^{0\nu}$ within the deformed QRPA  the tensor contribution has been neglected 
till now. Within the spherical QRPA this component of $M^{0\nu}$ reduces its total value  
by less than 10 \%~\cite{anatomy,rodin,src}. 

The single particle states are those of the axially symmetric Woods-Saxon  
mean field and are expressed in the basis of an axially-deformed   
harmonic oscillator states. The parametrization of the 
mean field is adopted from the spherical calculations  
of Refs. \cite{anatomy,rodin,src}. The single-particle model space  
consists of $4-6\hbar\omega$ shells in the spherical limit. 
Only the quadrupole deformation is taken into account in the  
calculation. The fitted values of the parameter $\beta_2$ of the deformed 
Woods-Saxon mean field, which allow us to reproduce the experimental $\beta_{B(E2)}$, 
are shown in Table \ref{table.1}. The spherical limit (i.e., $\beta_2=0$) 
is considered as well, to compare with the earlier results \cite{erice11} obtained in the spherical QRPA. 

We use the nuclear Brueckner G-matrix, obtained by a solution 
of the Bethe-Goldstone equation with the Bonn-CD one boson exchange  
nucleon-nucleon potential, as a residual interaction. The BCS equations 
are solved to obtain occupation amplitudes and quasiparticle energies, 
constituents of the nuclear Hamiltonian.  The pairing interactions  
are adjusted to fit the empirical pairing gaps for protons and neutrons. 
The gap parameters are determined phenomenologically from the  
odd-even mass differences through a symmetric five-term formula involving  
the experimental binding energies. The values obtained from this  
procedure for the nuclei under consideration can be seen in 
Table~\ref{table.1}. The calculated BCS overlap factors of the initial  
and final quasiparticle mean fields are given in the last  
column of Table~\ref{table.1}. This factor 
represents a possible suppression of 
the $0\nu$ECEC NME due to different  
deformations  of initial and final nuclei~\cite{deform}.   
 
To solve the deformed QRPA equations, one has to fix the particle-hole 
$g_{ph}$ and particle-particle $g_{pp}$ renormalization factors of the  
residual interaction of the nuclear Hamiltonian.  
Since experimental information about the position  
of the Gamow-Teller giant resonance for ${^{152}\mathrm{Gd}}$, ${^{164}\mathrm{Er}}$ and  
${^{180}\mathrm{W}}$ is not available we consider $g_{ph}=0.9$ like in previous 
calculation of the $0\nu\beta\beta$-decay NME of $^{150}$Nd and  
${^{160}\mathrm{Gd}}$. In~\cite{rodin} it was proposed to adjust the particle-particle strength parameter $g_{pp}$  
to the measured $2\nu\beta\beta$-decay half-lives, 
i.e., to reproduce the experimental value of the matrix element $M^{2\nu}_{GT}$. 
This procedure 
makes the $0\nu\beta\beta$-decay NMEs essentially independent  
of the size of the single-particle basis and the nuclear structure  
input. Due to a small $Q$-value, the half-life of the double-electron capture
with emission of two neutrinos ($2\nu$ECEC) of  
${^{152}\mathrm{Gd}}$, ${^{164}\mathrm{Er}}$ and ${^{180}\mathrm{W}}$ is too large to be measurable.  
Therefore, the parameter $g_{pp}$ of the QRPA  
is fixed by the assumption that the matrix element $M^{2\nu}_{GT}$ 
of the $2\nu$ECEC process lies within the range $(0,0.10)$~MeV$^{-1}$.  
Recall that  $M^{2\nu}_{GT}$ for double  
$\beta$-decaying nuclei from the region $^{128,130}$Te, $^{136}$Xe  
and $^{150}$Nd does not exceed  the above range by assuming  
the weak-axial coupling constant $g_A$ to be unquenched ($g_A = 1.269$)  
or quenched ($g_A = 1.0$). As it will be shown below this procedure 
of fixing $g_{pp}$ is not a significant source of uncertainty in 
the calculated $0\nu$ECEC half-lives.

\begin{table}[htb]
\centering
\caption{The matrix element $M^{0\nu}$ for the $2\nu$ECEC of 
${^{152}\mathrm{Gd}}$, ${^{164}\mathrm{Er}}$ and ${^{180}\mathrm{W}}$ calculated within the spherical 
and deformed QRPA with realistic nucleon-nucleon interaction (Bonn-CD potential) 
\cite{stone}.    
}\label{table.2}
\centering
\renewcommand{\arraystretch}{1.1} 
\begin{tabular}{lccccc}
\hline\hline 
Nucleus & $M^{2\nu}_{GT}$ & \multicolumn{4}{c}{ $M^{0\nu}$ } \\\cline{3-6}
        & [$MeV^{-1}$] & sph.  & & def. & def. \\ 
        &              & QRPA  & & QRPA ($\beta_2 =0$) & QRPA \\ 
\hline 
${^{152}\mathrm{Gd}}$ & 0.10 & 7.59 & & 7.50 & 3.23 \\
                                   & 0.00 & 7.21 & &      & 2.67 \\ 
${^{164}\mathrm{Er}}$ & 0.10 & 6.12 & & 7.20 & 2.64 \\ 
                                   & 0.00 & 5.94 & &      & 2.27 \\ 
 ${^{180}\mathrm{W}}$ & 0.10 & 5.79 & & 6.22 & 2.05 \\
                                   & 0.00 & 5.56 & &      & 1.79 \\ 
\hline
\hline
\end{tabular}
\end{table}

In Table \ref{table.2} the NME $M^{0\nu}$ for ${^{152}\mathrm{Gd}}$, ${^{164}\mathrm{Er}}$ and ${^{180}\mathrm{W}}$  calculated within the spherical and deformed QRPA    
are presented. There is a qualitative agreement between the results  
of the spherical QRPA and the spherical limit of the deformed QRPA ($\beta_2=0$).\footnote{Note, that $g_{ph}=0.9$ is used here in all calculations while the results of~\cite{erice11} were obtained with $g_{ph}=1.0$.} Since the deformed nuclei are described within the adiabatic Bohr-Mottelson approximation, the spherical limit of the deformed QRPA should be taken with caution. The differences can be attributed to the fact that within the spherical QRPA \cite{anatomy,src} one-body and two-body matrix elements entering the expressions for $M^{2\nu}_{GT}$ and $M^{0\nu}$, respectively, are calculated with single-particle wave functions approximated by the spherical harmonic oscillator ones, while realistic Woods-Saxon single-particle wave functions are used in the deformed QRPA \cite{dqrpa2,dqrpa3,dqrpa4}. In addition, $M^{0\nu}$ 
obtained within the spherical QRPA contains also $M_T$ contribution, which can reduce
its value by up to 10\%. The results in Table \ref{table.2} indicate that the nuclear 
deformation decreases the value of $M^{0\nu}$ by more than factor of 2-3. We note that 
the largest suppression of $M^{0\nu}$ due to deformation is realized for  $A=180$ nuclear system,  
where deformations of initial and final nuclei are comparable 
(same sign is assumed, see Table \ref{table.1}). 
It means that the suppression of $M^{0\nu}$ can be associated with the large deformation 
of initial and final nuclei and large value of A. Before the effect of deformation 
on $M^{0\nu}$ for nuclei with smaller A was associated with difference in  deformations 
of the initial and final nuclei \cite{deform,lssm,phfb}.   
 
\section{The $0\nu$ECEC half-lives of $^{152}\mathrm{Gd}$, $^{164}\mathrm{Er}$ and $^{180}\mathrm{W}$} 
 
Equations~(4) - (7) imply that the half-lives are determined by the following properties  
of the initial and final atoms:  
 
i) The mass difference $\Delta$ (\ref{e:6}) determines the proximity  
to the resonance condition and ultimately the magnitude of the effect.  
It depends on the $Q$-value and the energy of two electron vacancies in the final atom.  
The selection rule of Ref.~\cite{mik11} imply that it suffices to consider the capture of  $ns_{1/2}$ and $np_{1/2}$ electrons.  

\begin{table}[htb] 
\centering 
\caption{ 
Upper and lower average components of the Dirac bispinors $\langle f \rangle$ and $\langle g \rangle$  
in $^{158}$Gd and $^{166}$Er for $1s_{1/2}$, $2s_{1/2}$, $3s_{1/2}$, and $2p_{1/2}$ electron shells (in keV$^{3/2}$).  
The upper lines give solutions based on the Dirac equation in the Coulomb field \cite{mik11}, the lower lines give the same solutions based on the Dirac-Hartree-Fock method \cite{MAWA1973}. 
}\label{table.3} 
\begin{tabular}{lcc} 
\hline\hline  
Average  & $^{158}$Gd  & $^{166}$Er \\ 
\hline  
$\langle f(1s_{1/2}) \rangle$  &  $1.33 \times 10^{4}$ & $1.57 \times 10^{4}$            \\	 
                               &  $1.27 \times 10^{4}$ & $1.50 \times 10^{4}$           \\	 
$\langle f(2s_{1/2}) \rangle$  &  $5.20 \times 10^{3}$ & $6.20 \times 10^{3}$           \\	 
                              &         $4.59 \times 10^{3}$             &      $5.46 \times 10^{3}$             \\	 
$\langle f(3s_{1/2}) \rangle$  &         $2.84 \times 10^{3}$             &      $3.39 \times 10^{3}$             \\	 
                               &         $2.15 \times 10^{3}$             &      $2.60 \times 10^{3}$             \\	 
$\langle g(2p_{1/2}) \rangle$  &         $-1.12 \times 10^{3}$             &      $-1.43 \times 10^{3}$             \\	 
                               &         $-9.53 \times 10^{2}$             &      $-1.22 \times 10^{3}$             \\	 
\hline 
\hline 
\end{tabular} 
\end{table} 

Recently, $Q$-values in the $0\nu$ECEC transitions of $^{152}$Gd \cite{elis1}, $^{164}$Er \cite{elis4}, 
and $^{180}$W \cite{droese11} have been remeasured using Penning-trap mass spectrometry with an uncertainty of 180 eV, 120 eV, and 270 eV, respectively.
 
\begin{table*}[htb] 
\caption{ 
The calculated $0\nu$ECEC half-lives of $^{152}$Gd, $^{164}$Er, and $^{180}$W for $m_{\beta\beta} = 50 $ meV. 
The second and third columns show the quantum numbers of the electron holes.  
Here, $n$ is the principal quantum number, $j$ is the total angular momentum,  
and $l$ is the orbital momentum. Shown  
in the columns four, five and six are the hole energies and their Coulomb 
interaction energy (in units of keV). The column seven shows the radiative widths  
of the excited electron shells. The column eight shows the mass difference 
of the initial and final atoms. The last two columns show the minimum 
and maximum half-lives of the $0\nu$ECEC transitions (in years). 
The masses and  the energies are in keV. 
}\label{table.4}    
\centering 
\renewcommand{\arraystretch}{1.1}  
\begin{tabular}{cccrrrccccc} 
\hline \hline 
Nucleus &$(n 2j l)_a$ & $(n 2j l)_b$ & $E_{a}\;\;$ & $E_{b}\;\;$ & $E_{C}$ & $\Gamma_{ab}$ (keV)&  
$\Delta$ (keV) & & $T^{\min}_{1/2}$ (y) & $T_{1/2}^{\max}$ (y) \\ 
\hline 
$^{152}$Gd  
  & 110 & 210 & 46.83 & 7.74 & 0.34   & $2.3 \times 10^{-2}$ & $ -0.83 \pm 0.18$ & & $4.7 \times 10^{28}$ & $4.8 \times 10^{29}$  \\ 
  & 110 & 211 & 46.83 & 7.31 & 0.32   & $2.3 \times 10^{-2}$ & $ -1.27 \pm 0.18$ & & $4.2 \times 10^{31}$ & $1.1 \times 10^{32}$  \\ 
  & 110 & 310 & 46.83 & 1.72 & 0.11   & $3.2 \times 10^{-2}$ & $ -7.07 \pm 0.18$ & & $9.4 \times 10^{31}$ & $1.1 \times 10^{32}$  \\ 
$^{164}$Er  
  & 210 & 210 &  9.05 & 9.05 & 0.22   & $8.6 \times 10^{-3}$ & $ - 6.82 \pm 0.12$ & & $7.5 \times 10^{32}$ & $8.4 \times 10^{32}$ \\  
  & 210 & 211 &  9.05 & 8.58 & 0.23   & $8.3 \times 10^{-3}$ & $ - 7.28 \pm 0.12$ & & $4.2 \times 10^{34}$ & $4.6 \times 10^{34}$ \\  
  & 210 & 310 &  9.05 & 2.05 & 0.11   & $1.8 \times 10^{-2}$ & $ -13.92 \pm 0.12$ & & $3.5 \times 10^{33}$ & $3.9 \times 10^{33}$ \\  
$^{180}$W  
  & 110 & 110 &  63.35 & 63.35 & 1.26 & $7.2 \times 10^{-2}$ & $-11.24 \pm 0.27$ & & $1.3 \times 10^{ 31}$ & $1.8 \times 10^{31}$ \\  
\hline \hline  
\end{tabular}  
\end{table*}    
 
In atomic physics, the electron binding energies are usually measured to within a few eV.  
We used the binding energies of single electron holes $ E_{a}$ from Ref. \cite{larkins}.  
Noticeable corrections arise from the Coulomb interaction between two holes $E_C$.  
Coulomb interaction energy is calculated on the basis of the Dirac equation with account  
of the screening of the nuclear charge, which gives an accuracy comparable to the present  
experimental errors in $Q$-values. More accurate estimates can be obtained by averaging the  
Fermi-Breit potential over the states of the two vacancies.

ii) In the unitary limit the capture rate is inversely proportional to the width of the decay of excited atoms  
in the final state. The radiative width of the electron shell with two holes is estimated as  
$ \Gamma_{ab} = \Gamma_{a} + \Gamma_{b} $ on the basis of the measured and recommended values  
of radiative widths $ \Gamma_{a} $ of the single vacancies \cite{campbel}.  
The radiative width also determines the desired accuracy in measuring $Q$-values. 
 
iii) In nuclei with large values of $Z$, electrons in the lower shells are moving at a speed close  
to the speed of light. The wave functions of electrons in the Coulomb field plus the self-consistent  
field of neighboring electrons must be considered in a relativistic approach.  
The upper and lower radial  
functions of the Dirac bispinors are averaged over the volume of the nucleus. Relativistic effects lead to  
an increase in the probability of $ 0\nu$ECEC process. Estimates of the average values of the radial  
component of the Dirac wave functions inside the nucleus can be obtained using well-known wave functions  
of electrons in the Coulomb field,  
or by solving many-body problem using the Dirac-Hartree-Fock method. 
The $ns_{1/2}$ electron capture is proportional to the mean value of the upper bispinor radial component,  
while the capture from the $np_{1/2}$ states is determined by the mean value of  
the lower radial components of the Dirac wave function. In Table \ref{table.3}  
the average values of the upper and lower radial components of Ref. \cite{mik11}  
are compared with the values obtained in the Dirac-Hartree-Fock method \cite{MAWA1973} 
for $^{158}$Gd and $^{166}$Er. The agreement is quite good.  
 
iv) The probability of capture is proportional to the square of the nuclear matrix element,  
which we discussed in the previous section in detail. 
 
In Table \ref{table.4} the maximum and minimum values of the $0\nu$ECEC half-lives (in years)
of ${^{152}\mathrm{Gd}}$, ${^{164}\mathrm{Er}}$ and $^{180}\mathrm{W}$  
for $m_{\beta\beta} = 50$ meV assuming the capture of most favored atomic electrons, 
the associated values of the binding energies and  
the Coulomb interaction energy of the two holes, and 
the mass difference $\Delta$ of initial and final excited atoms are shown. 
The estimates for $T^{\min}_{1/2}$ are taken for the minimum mass  
difference deviating by not more than three standard  
errors from the experimental mean value unlike in \cite{erice11}, where the unitary limit was considered. 
Among the transitions, the favoured one is 
the capture of electrons from K and L shells in the case of $^{152}\mathrm{Gd}$,  
which results in the half-life in the range $4.7\times 10^{28} \div 4.8\times 10^{29}$ years.  
This transition is still rather far from the resonant level. The half-life appears thereby 2 - 3 orders of magnitude greater  
as compared to the half-life of $0\nu\beta\beta$ decay of $^{76}$Ge \cite{src}.  
 
\section{Conclusion} 
 
As shown in Ref.~\cite{mik11}, the $0\nu$ECEC half-lives of a dozen of isotopes are comparable  
to the shortest half-lives of the $0\nu \beta \beta$ decays of nuclei provided  
the resonance condition is matched with an accuracy of tens of electron-volts.  
Among the promising isotopes $^{152} $Gd, $ ^{164} $Er, and $^{180} $W  
were found to be associated with the transitions between ground states of the nuclei.  
The estimates of the $0\nu$ECEC half-lives were recently improved by more accurate  
measurements of $Q$-values for $^{152} $Gd \cite{elis1}, $ ^{164} $Er \cite{elis4}, 
and $^{180} $W \cite{droese11} in a Penning trap.  

In this paper, we have made a further step to refine the estimates of the half-lives by going 
beyond the spherical approximation in the calculation of nuclear matrix elements.  
We found within the deformed QRPA that the deformation of the nuclei leads to  
suppression of the NMEs by the factor of 2-3 as compared with the spherical limit.  
The suppression of NMEs depends not only on the relative deformation of the initial  
and final nuclei, but also on their absolute values.  
 
We conclude that the $0\nu$ECEC half-life of $^{152}$Gd is 2-3 orders of magnitude longer than 
the half-life of $0\nu\beta\beta$ decay of $^{76}$Ge corresponding to the same value of 
the Majorana neutrino mass. Our calculation excludes $ ^{164} $Er and $ ^{180} $W from the 
list of prospective isotopes to search for the neutrinoless double-electron capture. 
 
\acknowledgments 
 
This work is supported in part by the Deutsche Forschungsgemeinschaft  
within the projects SFB TR27 "Neutrinos and Beyond", 436 SLK 17/298 and 436 RUS 113/721/0-3. F.~\v{S}. acknowledges support by the VEGA Grant agency  
of the Slovak Republic under the contract No.~1/0639/09. M.I.K. is partly supported
by Grant No. 09-02-91341 of the Russian Foundation for Basic Research. 
 


\begin{thebibliography}{99} 
 
\bibitem{seesaw} M.~Gell-Mann, P.~Ramond, and R.~Slansky, in {\it Supergravity}, 
  p.~315, edited by F. van Nieuwenhuizen and D. Freedman, North Holland, Amsterdam, 1979\,; 
  T.~Yanagida, Proc. of the {\it Workshop on Unified Theory and the Baryon Number of the 
  Universe}, KEK, Japan, 1979\,; 
  R.N. Mohapatra and G.~Senjanovi{\'c}, Phys. Rev. Lett. {\bf 44}, 912 (1980); 
  P. Minkovski, Phys. Lett. B {\bf 67}, 421 (1977). 
%
%
%
\bibitem{dbdrev} F.T. Avignone, S.R. Elliott and J. Engel, Rev. Mod. Phys. {\bf 80}, 481 (2008).  
 
\bibitem{DERU} J. Bernab\'eu, A. De Rujula, and C. Jarlskog, Nucl. Phys. B {\bf 223}, 15 (1983). 

\bibitem{AME2003} G. Audi, O. Bersillon, J. Blachot, A.H. Wapstra, Nucl. Phys. A \textbf{729}, 3 (2003).
 
%
%
\bibitem{penning1} G. Douysset, T. Fritioff, C. Carlberg, I. Bergstrom, M. Bjorkhage,
  Phys. Rev. Lett. {\bf 86}, 4259 (2001).  

\bibitem{penning2} K. Blaum, Phys. Rep. {\bf 425}, 1  (2006). 
\bibitem{penning3} K. Blaum, Yu. N. Novikov, and G. Werth, Contemp. Phys. {\bf 51}, 149 (2010). 
 
\bibitem{sujwy}  Z. Sujkowski and S. Wycech, Phys. Rev. C {\bf 70}, 052501 (2004); 
 
\bibitem{lukas}  L. Lukaszuk, Z. Sujkowski and S. Wycech, Eur. Phys. J A {\bf 27}, 63 (2006). 
 
\bibitem{SIM08} F. \v Simkovic, M.I. Krivoruchenko, Phys. Part. Nucl. Lett. {\bf 6}, 298 (2009). 
 
\bibitem{mik11} M.I. Krivoruchenko, F. \v Simkovic, D. Frekers, A. Faessler, 
   Nucl. Phys. A {\bf 859}, 140 (2011). 
 
\bibitem{elis3}   S. Eliseev \textit{et al.}, Phys. Rev. C {\bf 84}, 012501 (2011).  

\bibitem{erice11} 
  F. \v Simkovic, M.I. Krivoruchenko, A. Faessler, 
  Prog. Part. Nucl. Phys. {\bf 66}, 446 (2011).   

\bibitem{verg11} J. D. Vergados, Phys. Rev. C {\bf 84}, 044328 (2011). 

\bibitem{redshaw09} 
 M. Redshaw, E. Wingfield, J. McDaniel, E.G. Myers, Phys. Rev. Lett. {\bf 98}, 053003 (2007);
 M. Redshaw, B.J. Mount, E.G. Myers, F.T. Avignone, Phys. Rev. Lett. {\bf 102}, 212502 (2009).  
 
\bibitem{SCIE09} N.D. Scielzo \textit{et al.}, Phys. Rev. C {\bf 80}, 025501 (2009).  
 
\bibitem{RAKH09} S. Rahaman \textit{et al.}, Phys. Rev. Lett. {\bf 103}, 042501 (2009).  
 
\bibitem{kolhinen} V.S, Kolhinen \textit{et al.}, Phys. Lett B {\bf 684}, 17 (2010).  
 
\bibitem{mount10} B. J. Mount, M. Redshaw, and E. G. Myers, Phys. Rev. C  
  {\bf 81}, 032501 (2010). 
 
\bibitem{elis1} S. Eliseev \textit{et al.}, Phys. Rev. Lett. {\bf 106}, 052504 (2011).  
 
\bibitem{elis2}   S. Eliseev \textit{et al.}, Phys. Rev. C {\bf 83}, 038501 (2011).  
  
\bibitem{elis4}   M. Goncharov \textit{et al.}, Phys. Rev. C {\bf 84}, 028501 (2011).  
 
\bibitem{droese11} Ch. Droese \textit{et al.}, accepted in Nucl. Phys. A (2011),
 	arXiv:1111.6377v1 [nucl-ex].

%
%
 
\bibitem{barab1}  A.S. Barabash, Ph. Hubert, A. Nachab, and V. Umatov,  
   Nucl. Phys. A {\bf 785}, 371 (2007). 
 
\bibitem{barab2}  A.S. Barabash, Ph. Hubert, A. Nachab, S.I. Konovalov,  
 I.A. Vanyushin, and V.I. Umatov, Nucl. Phys. A {\bf 807}, 269 (2008). 
 
\bibitem{belli09} P. Belli \textit{et al.}, Nucl. Phys. A {\bf 842}, 101 (2009).  
 
\bibitem{rukh11} N.I. Rukhadze \textit{et al.}, Nucl. Phys. A {\bf 852}, 197 (2011). 
 
\bibitem{frek11} D. Frekers \textit{et al.}, Nucl. Phys. A {\bf 860}, 1 (2011). 
 
\bibitem{belli11} P. Belli \textit{et al.}, Eur. Phys. J. A {\bf 47}, 91 (2011). 

\bibitem{belli11a} P. Belli \textit{et al.}, arXiv: 1110.3690 [nucl-ex]. 
 
\bibitem{suho2011} J. Suhonen, Phys. Lett. B \textbf{701}, 490  (2011). 
 
\bibitem{MAWA1973} J. B. Mann and J.T. Waber, Atomic Data {\bf 5}, 201 (1973).  
 
%
%
 
\bibitem{dqrpa1} M. Saleh Yousef, V. Rodin, A. Faessler, F. \v Simkovic,  
    Phys. Rev. C {\bf 79}, 014314 (2009). 
 
\bibitem{dqrpa2} D. Fang, A. Faessler, V. Rodin, M. Yousef Saleh, F. \v Simkovic, 
    Phys. Rev. C {\bf 81}, 037303 (2010).  
 
\bibitem{dqrpa3} D. Fang, A. Faessler, V. Rodin, F. \v Simkovic,  
    Phys. Rev. C {\bf 82}, 051301 (2010). 

\bibitem{dqrpa4} D. Fang, A. Faessler, V. Rodin, F. \v Simkovic,
  Phys. Rev. C {\bf 83}, 034320 (2011).
%
%
 
\bibitem{BPont} B. Pontecorvo, J. Exptl. Theoret. Phys. {\bf 33}, 549 (1957) 
  [Sov. Phys. JETP {\bf 6}, 429  (1958)]; J. Exptl. Theoret. Phys. 
  {\bf 34}, 247  (1958) [Sov. Phys. JETP {\bf 7}, 172 (1958)]. 
 
\bibitem{MNS} Z. Maki, M. Nakagawa, and S. Sakata, Prog. Theor. Phys. {\bf 28}, 870 (1962). 
 
%
%
%
 
\bibitem{Sim99} F. \v{S}imkovic, G. Pantis, J. D. Vergados, and A. Faessler,  
Phys. Rev. C{\bf 60}, 055502 (1999). 
 
\bibitem{anatomy} F. \v Simkovic, A. Faessler, V.A. Rodin, P. Vogel, and 
  J. Engel, Phys. Rev. C {\bf 77}, 045503 (2008). 
%
%
\bibitem{stone} N.J. Stone, At. Data Nucl. Data Tables {\bf 90}, 75 (2005);
see also ``Chart of nucleus shape and size parameters", \url{http://cdfe.sinp.msu.ru/services/radchart/radmain.html} 
\bibitem{raman} S. Raman, C. H. Malarkey, W. T. Milner, C. W. Nestor, Jr., and 
     P. H. Stelson, At. Data Nucl. Data Tables {\bf 36}, 1 (1987). 
 
\bibitem{deform} F. \v Simkovic, L. Pacearescu, A. Faessler, 
  Nucl. Phys. A {\bf 733}, 321 (2004);  
 R. Alvarez-Rodriguez \textit{et al.}, Phys. Rev. C {\bf 70}, 064309 (2004). 
%
%
\bibitem{rodin} V.A. Rodin, A. Faessler, F. \v Simkovic and P.~Vogel, 
   Phys. Rev. C, {\bf 68}, 044302 (2003); Nucl. Phys. A {\bf 107}, 
   (2006); {\bf 793}, 213(E) (2007). 
\bibitem{src}  
 F. \v Simkovic, A. Faessler, H. M\"uther, V. Rodin, M. Stauf, Phys. Rev. C  
 {\bf 79}, 055501 (2009). 
  
\bibitem{lssm} J. Men\'endez, A. Poves, E. Caurier, and F. Nowacki,  
    Nucl. Phys. A {\bf 818}, 139 (2009). 
 
\bibitem{phfb} R. Chandra, J. Singh, P. K. Rath, P. K. Raina, and J. G. Hirsch, 
    Eur. Phys. J. A {\bf 23}, 223 (2005); S. Singh, R. Chandra, P. K. Rath, 
    P. K. Raina, and J. G. Hirsch, ibid. 33, {\bf 375} (2007); P.K. Rath,  
    R. Chandra, K. Chaturvedi, P.K. Raina, J.G. Hirsch, 
    Phys. Rev. C {\bf 82},   064310 (2010). 
 
 
\bibitem{larkins} F.B. Larkins, At. Data Nucl. Data Tables {\bf 20}, 313 (1977). 
 
\bibitem{campbel} J.L. Campbell and T. Papp,  
     At. Data Nucl. Data Tables {\bf 77}, 1 (2001). 
 
\end{thebibliography}
\end{document}